\documentclass[12pt]{article}
\usepackage{newtxtext,newtxmath}

\usepackage{graphicx}

\usepackage[letterpaper,margin=1in]{geometry}

\renewenvironment{abstract}
	{\quotation}
	{\endquotation}

\date{}

\makeatletter
\renewcommand{\fnum@figure}{\textbf{Figure \thefigure}}
\renewcommand{\fnum@table}{\textbf{Table \thetable}}
\makeatother

\usepackage{scicite}

\usepackage{url}

\def\scititle{Exploring Biological Neuronal Correlations with Quantum Generative Models
}
\title{\bfseries \boldmath \scititle}

\author{
	Vinicius~Hernandes$^{\ast}$,
	Eliska~Greplova\and
 \small  QuTech and Kavli Institute of Nanoscience, Delft University of Technology, Delft, the Netherlands.\and
	\small$^\ast$Corresponding author. Email: v.hernandes@tudelft.nl\and
}

\begin{document} 

\maketitle


\begin{abstract} \bfseries \boldmath

Understanding of how biological neural networks process information is one of the biggest open scientific questions of our time. Advances in machine learning and artificial neural networks have enabled the modeling of neuronal behavior, but classical models often require a large number of parameters, complicating interpretability. Quantum computing offers an alternative approach through quantum machine learning, which can achieve efficient training with fewer parameters. In this work, we introduce a quantum generative model framework for generating synthetic data that captures the spatial and temporal correlations of biological neuronal activity. Our model demonstrates the ability to achieve reliable outcomes with fewer trainable parameters compared to classical methods. These findings highlight the potential of quantum generative models to provide new tools for modeling and understanding neuronal behavior, offering a promising avenue for future research in neuroscience.

\end{abstract}

\noindent Exploring the information processing within biological neuronal networks remains a core challenge in contemporary science, with direct implications across disciplines like neuroscience, medicine, and deep learning \cite{RiWaRu1999, WhWiHe2010, HaKuSu2017,
SaNeSu2021}. One way to approach this problem is to use computational models that can reproduce the neuronal activity data produced in real systems. Accurate synthetic data can be extremely useful to study properties such as network connectivity and response to stimuli under controlled conditions \cite{WePeBr2022, SeKrDj2022}. 

Several models for neuronal activity have been developed, and many achieve outstanding results in replicating neuronal network correlations. One class of methods that use statistical mechanics tools to model neuronal activity are Maximum Entropy models, which reliably capture some network correlations by only fitting pairwise interactions \cite{TkScBe2006, ScBeSe2006}. Even though numerous adaptations of this technique have been implemented to achieve higher accuracy or to include temporal correlations \cite{TaJaHo2008, MaBoFr2009, TkMaMo2013, GrTkSe2013, TkMaAm2014, DelamareFerrari2022}, this approach shows several limitations when addressing larger networks, especially due to its assumption that pairwise correlations are sufficient to encapsulate most of the statistical features of these complex systems \cite{RoNiLa2009, KoSoGr2014}.

Another effective approach for modeling neuronal activity is to use machine learning (ML) models to produce data that fits the biological network statistics and use the model to further investigate the properties of the real system. The ML models do not rely on prior information about the biological system but instead learn to reproduce correlations solely from data. A supervised strategy using Convolutional Neural Networks first showed that a deep learning approach can be successful at generating neural responses from stimuli \cite{McMaNa2016}. However, the model's benefits are hampered by limited accuracy and its dependency on labeled data. With the increasing popularization of generative models, models with superior predictive performance and generalization power were implemented. Models like Variational Auto-Encoders \cite{PaOsCo2018}, Recurrent Neural Networks \cite{BeWaMo2021}, Generative Adversarial Networks (GANs) \cite{MoOnPi2018}, and Transformers \cite{LeShlizerman2022} have been used to produce spike trains (binary sequence representing neuronal activity) with high accuracy and good correspondence of spatial and temporal correlations when compared to real data. While each iteration of these models improves in quality, all share the same disadvantage regarding their interpretability. In order to fit the statistics of larger systems, these models need to use a number of trainable parameters that scales unfavourably with the number of simulated neurons. Apart from demanding more computational power, excessive number of parameters make the models difficult to analyze or be used as a tool to investigate concrete properties of biological networks.

As the field of quantum computing rapidly advances, quantum machine learning (QML) models are rising as an alternative to classical methods, with the possibility of achieving similar results while keeping the parametrized model more compact in terms of trainable parameters \cite{Schuld2018, ScBoSv2020, DunjkoWitted2020, AbSuZo2021}. Specifically, the field of quantum generative learning received much attention recently: quantum models have shown better generalization and expressivity for specific tasks when compared to their classical counterparts \cite{BeGaPe2019, DuHsMi2020, DuTuWu2022}. Since the conception of QML, one class of quantum generative models has been extensively studied: quantum generative adversarial networks (QGANs) \cite{DallaireKilloran2018}. The adversarial approach has proven successful and is being continually improved, producing higher-dimensional data with more stable training routines \cite{ZoLuWo2019, HaZhYu2020, HuDuYu2021, TsWeEr2023, ZhZhTi2023}. 

This work is inspired by the observation that quantum generative models have shown promise in replication of discrete distributions \cite{TsWeEr2023, ChHuMa2023}. Additionally, the salamander retina dataset has been used as a benchmark for distribution learning using quantum Boltzmann machines \cite{kappen2020learning, HuCoNa2024}. These observations suggest a possibility of full reconstruction of both spatial and temporal correlations with a quantum generative model that we present here. We build on our preliminary work \cite{hernandes2023modeling} and design SpiQGAN, an efficient quantum framework that enables the production of synthetic neuronal data for biological neuronal networks. SpiQGAN generates spike trains of neuronal activity: data that consist of binary activation states of the neurons obtained from recording the response of ganglion cells of the salamander retina to a visual stimulus as a function of time \cite{marre2017multi}. This data set represents one of the standard benchmarks in neuronal activation modelling.

To achieve generation of the data that maximally resembles the real biological sample, we apply a hybrid quantum generative adversarial network, with a quantum generator that produces synthetic activity data, and a classical critic that aims to distinguish real data from the dataset \cite{marre2017multi} from those produced by the quantum generator. The model is trained adversarially, and the outcome is a generator that can reproduce neuronal activity that is to the high degree similar to the salamander retina dataset. Compared to classical neural networks alternatives, the quantum generator has the advantage of achieving reliable outcomes with a significantly reduced number of trainable parameters, that scale more favourably for increasing systems' sizes: the number of paramaters is linear in the number of neurons. In other words, SpiQGAN is able to reproduce the behavior of this  complex neuronal data set in both space and time with significantly fewer trainable parameters than classical ML models, thus forming a stepping stone towards using quantum approaches for more compact and more interpretable models for neuronal behavior.

\subsection*{Quantum Generative Model}

GANs are composed of two networks, the generator, whose goal is to generate data indistinguishable from the real dataset, and the discriminator, which tries to classify if a sample comes from the real dataset, or if is produced by the generator \cite{goodfellow2014generative, goodfellow2016nips, salimans2016improved}. The generator works by mapping a noise vector $\boldsymbol{z}$, sampled from a prior distribution $P_{\boldsymbol{z}}$, to an output $G(\boldsymbol{z})$. The discriminator takes as input either a real sample $\boldsymbol{x}$, sampled from the real distribution $P_{\boldsymbol{x}}$, or a fake sample $G(\boldsymbol{z})$ produced by the generator. The two networks are then trained adversarially so that the generator produces increasingly more realistic samples and the discriminator gets continuously better at discerning between real and fake samples.

The quantum version of GAN can contain a quantum generator, a quantum discriminator, or both. Making one of the components quantum means replacing the neural network with a parametrized quantum circuit (PQC). 

In most cases, QGANs are implemented using a quantum generator and a classical discriminator, focusing the use of a quantum circuit uniquely for the computationally heavier task of a GAN: generating samples by fitting the data's distribution. Huang et al. \cite{HuDuYu2021} introduced a resource-efficient QGAN for image generation, in which fixed PQC ansatz - with different parameters - is reused to produce different sections of output. This approach makes QGAN with a limited number of qubits a powerful tool to produce high-dimensional data. As with their classical counterpart, previous implementations of QGANs utilize the standard optimization routine, with a discriminator classifying samples. More recently, the so-called Wasserstein QGANs rose to prominence by replacing the discriminator with a critic and showing more stable training \cite{chakrabarti2019quantum, HeObRo2021}.

\begin{figure} 
    \centering
   \includegraphics[width=\textwidth]{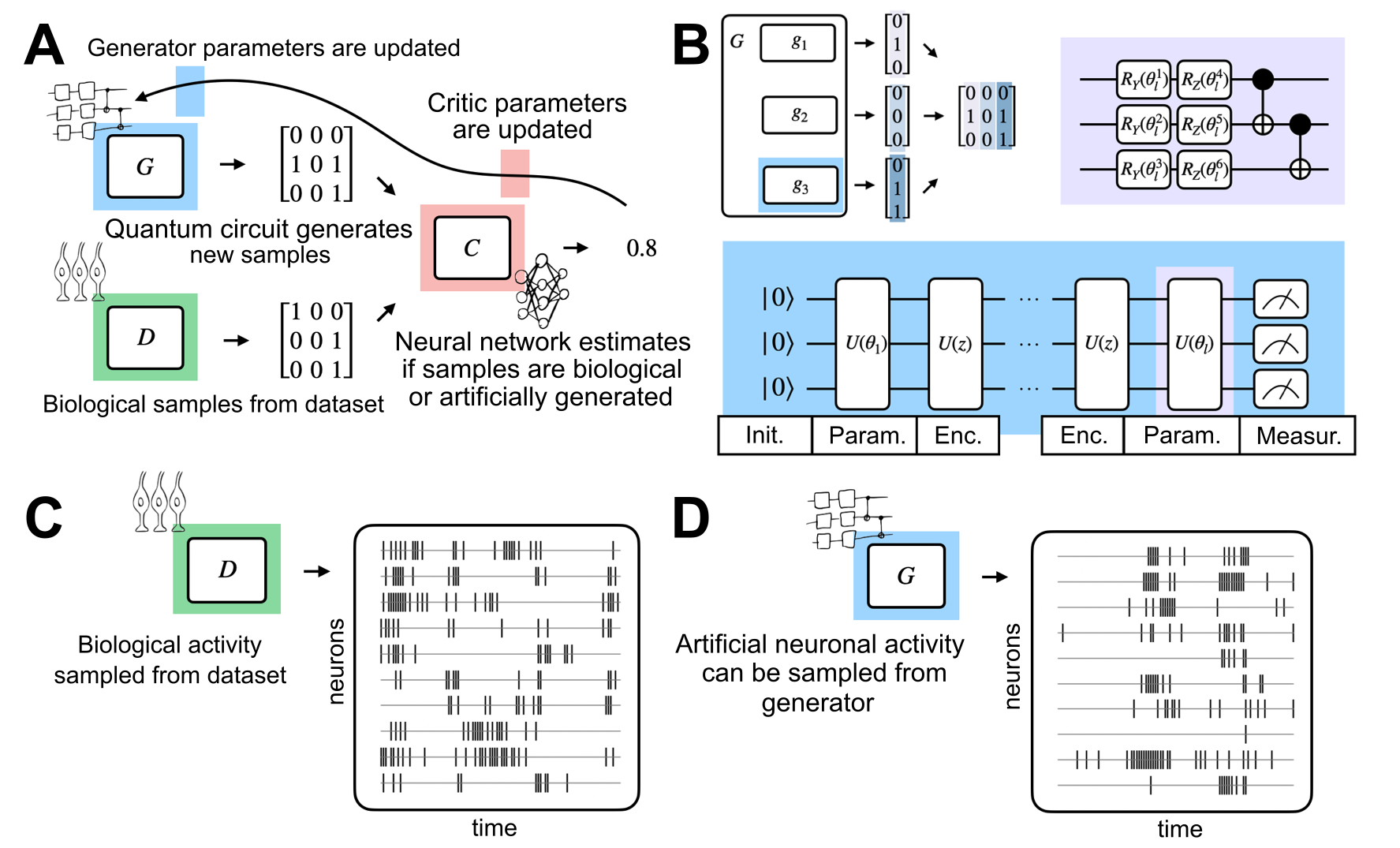} 

	\caption{\textbf{Illustration of the model architecture.} (A) Architecture of the model, with generator G producing generated samples, and dataset D producing biological samples, which are both used as input for critic C. (B) Architecture of generator. In the upper left corner, the generator composed of several sub-generators is shown. The bottom part shows that each sub-generator is a quantum circuit following a re-uploading scheme. Here a noise-encoding layer and a parametrized layer are repeated for $l$ layers, with the parametrized layer ansatz of each parametrized layer shown in the top right side. After trained, the generator can be used to produce samples (D) similar to samples obtained from the biological dataset (C).}
	\label{fig:circuit} 
\end{figure}

SpiQGAN uses the quantum generator circuit to produce samples of neuronal activity for $n$ neurons and $t$ timesteps. Specifically, we deploy a Patch Wasserstein QGAN scheme first introduced in quantum setting in Ref.~\cite{TsWeEr2023}. The overall training routine is shown in Fig.~\ref{fig:circuit}(A), where the generator produces samples that are fed to the critic, which estimates a distance between the distribution of data produced by the generator and that of the biological dataset. Fig.~\ref{fig:circuit}(B) shows the architecture of the generator, composed of $t$ sub-generators $g_i$, where each sub-generator uses the same parametrized quantum circuit ansatz with a different set of trainable parameters. For the generation of spike trains, each sub-generator with qubits is responsible for producing the activity state of $n$ neurons for one timestep. The result of each sub-generator is then concatenated to obtain the activity of $n$ neurons for $t$ timesteps. 

The PQC of the sub-generators is composed of $a$ auxiliary and $n$ feature qubits. Here, the task of producing fake samples from noise translates to mapping a random initial state $\lvert\boldsymbol{z}\rangle$ to a final state $\rvert\textit{g}\rangle$, followed by sampling one bit-string by measuring all feature qubits. The mapping is performed using a data re-uploading scheme \cite{PeCeGi2020, ScSwMe2021}. In the re-uploading scheme parametrized unitaries $U(\theta_i)$ are alternated with noise encoding unitaries $U(z)$, for $l = 5$ repeating layers, changing the parameters $\theta_i$ in each layer - Fig. \ref{fig:circuit}(B). In this setting, the encoder unitary consists of applying $R_X(\gamma)$ gates to all qubits, where $\gamma$ being a random angle uniformly sampled from the interval $[0, \pi]$. In the parametrized unitary, $U(\theta_i)$, a sequence of $R_Y({\theta}_i^k)$, $R_Z({\theta}_i^{k+1})$ gates are applied to every qubit, followed by $\mathrm{CNOT}$ gates between each pair of nearest-neighboring qubits.
The final state of the sub-generator is $\lvert g \rangle= \left(\Pi_i^l U(\theta_i)U(z)\right) \lvert 0 \rangle^{\otimes n}$. The patch output is obtained by measuring the feature qubits in the computational basis.

The output of the generator is produced by concatenating each sub-generator patch. In our representation, concatenating the patches is equivalent to stacking timesteps of the spike train. The critic consists of a fully-connected network with an input size equal to that of a sample taken from the generator or from the real data, one hidden layer with 64 units, with a ReLU activation function, and one output layer with no activation function, representing the distance between the probability distribution of the real and fake data.

SpiQGAN is trained using the loss functions

\begin{equation}
    \mathcal{L}_C = \frac{1}{2B} \sum_j C(G(\boldsymbol{z_j})) - C(\boldsymbol{x_j}),
    \label{eq:loss_C}
\end{equation}
and
\begin{equation}
    \mathcal{L}_G = - \frac{1}{B} \sum_j C(G(\boldsymbol{z_j})) - K\left(\sum_i G({z_j})^i-x_j^i\right),
    \label{eq:loss_G}
\end{equation}
for the critic and the generator, respectively. An important modification we made here is the addition of a biologically informed term in the generator loss function, $K(\sum_i G({z_j})^i-x_j^i)$, that is inspired by Maximum-Entropy models \cite{TkMaMo2013} and corresponds to the difference between the number of spikes in a fake sample and those in a real sample. We tested two variants of models, using a biologically-informed \textit{K}-loss function, with $K=1$ in Eq.~\ref{eq:loss_G}, and a standard (non-biologically inspired) loss function, with $K=0$. The normalization factor, $B$, corresponds to the batch size, equal to $32$ in this work. We use $\mathrm{Adam}$ as optimizer with learning rates equal to $0.05$ for the generator and $0.002$ for the critic. The dataset used to train the model is the neuronal activity recorded from retinal ganglion cells of the salamander retina (38). The goal of SpiQGAN is to produce a binary signal as a function of time for every simulated neuron. More specifically, we want first and second order correlations, like mean firing rates and covariance between neurons, respectively, of a generated sample to match those of a real sample.

\subsection*{Results}

We trained SpiQGAN for for $t = \{1, 5, 10, 20, 30\}$ timesteps and for $n = \{2, 4, 6, 8, 10\}$ neurons, in order to evaluate the quality of the generated data as a function of system size and time trace length.

Comparing the distribution of possible states of the simulated data to that of the salamander retina data is a straightforward way to evaluate the quality of our generative model. Reconstruction of these distributions also allows us to calculate distributions distances such as the the Jensen-Shannon (JS) divergence. However, direct distribution comparison is particularly challenging: for $n$ neurons and $t$ time steps, the number of possible (spiking) states is $2^{nt}$. We are thus able to directly visually compare distributions only for a small number of neurons. 

For two neurons, $n = 2$, and and one time step, $t = 1$, the possible states are $\{00, 01, 10, 11\}$. For two neurons and two time steps, $t = 2$, the possible states are $\{0000, 0001, ..., 1110, 1111\}$. In this representation, the first two bits represent neurons $1$ and $2$ at time step $1$ and the last two the states of these neurons at time step $2$. This means that the distribution of states quickly becomes intractable for increasing number of neurons or time steps. Nonetheless, it is very informative to compare distributions directly for low number of neurons.

For all cases, regardless of system size, we calculated a series of statistical values useful to evaluate the behavior of the generated and the real data from the salamander retina dataset. Specifically, we calculate the pairwise covariance between the activation state of a pair of neurons; the mean firing rate, which correspondent to how many times a neuron spikes per second; the k-probability, equal to the probability of k neurons being active at the same time; and the autocorrelogram to estimate the correlation between a trace of spikes and itself for delayed timesteps \cite{methods} .

\begin{figure} 
    \centering
   \includegraphics[width=\textwidth]{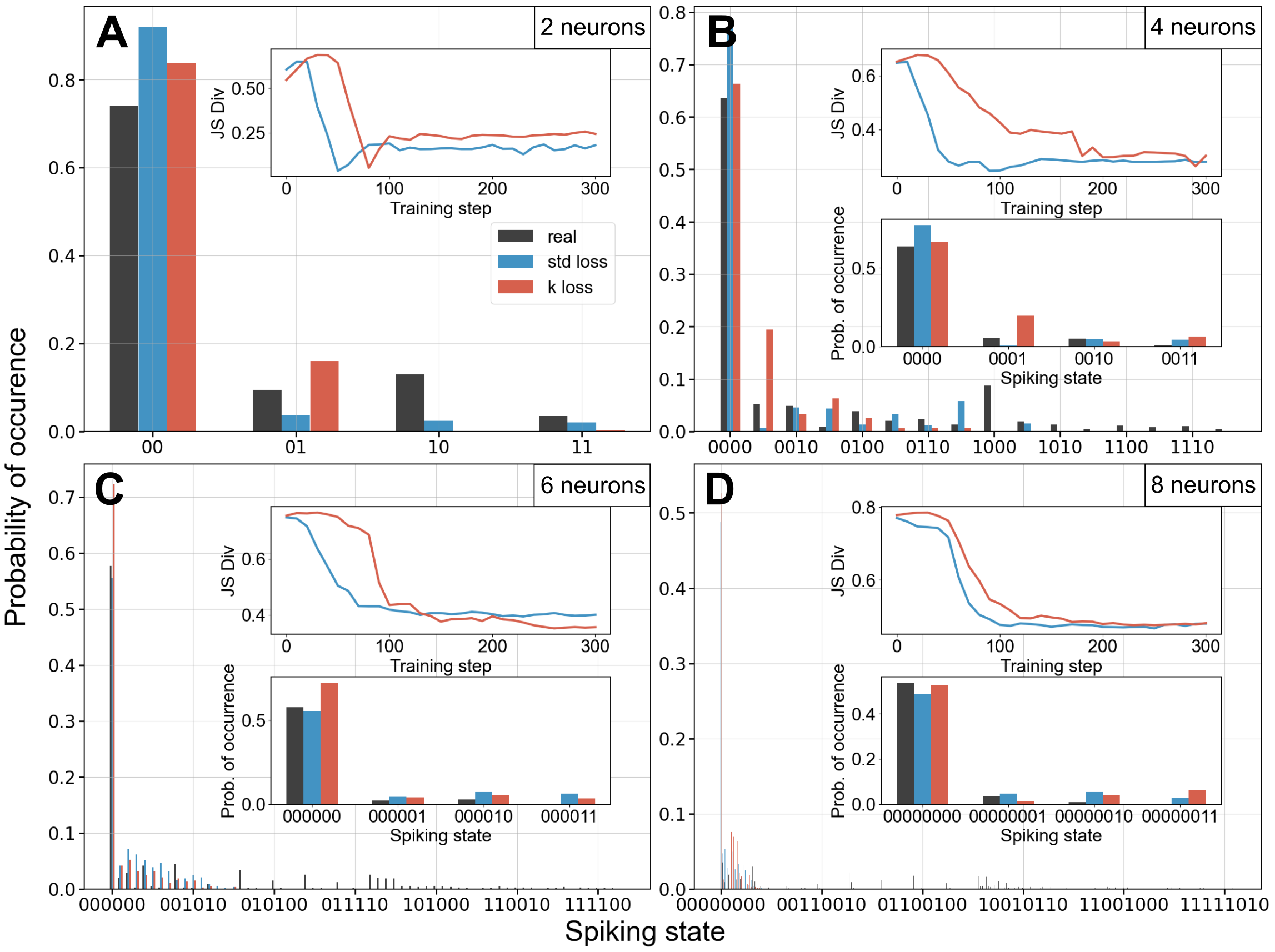} 

	\caption{\textbf{Comparison between distribution of states and JS divergence calculated using generated and real data.}
		Each panel show the distribution of spiking states for generated data obtained after training with the K-loss (in red) and with the standard loss (in blue), and the real distribution of the spiking states (black), for (A) 2, (B) 4, (C) 6, and (D) 8 neurons, all for the case of 1 timestep. The bottom inset shows a zoom of the first four activation states. The upper inset shows the JS divergence for all training steps, for K (red) and standard (blue) loss.}
	\label{fig:histograms} 
\end{figure}

First we consider the case with a unique timestep (one subgenerator quantum circuit), and neurons varying from 2 to 8. In these cases the distribution is easily numerically tractable. We show the final distribution of generated spiking states, compared to the distribution of the real data in Fig.~\ref{fig:histograms}, with a zoom on the most prominent terms of the distribution in the bottom inset. The probabilities of the spiking states are calculated using the last iteration of the trained circuit. This is coincident with the last value of the JS divergence, which steadily decreases during training, visible in the bottom insets of Fig.~\ref{fig:histograms}. These results show that for a sufficient number of training steps the distribution of generated states converges to the distribution of the salamander retina dataset. This distribution convergence is a first indication that the training is working as intended, and the samples produced by the generator match some of the statistics of the real data. Throughout, we compared both standard loss and biologically inspired \textit{K}-loss. We found that on average \textit{K}-loss performed slightly better.

\begin{figure} 
    \centering
   \includegraphics[width=\textwidth]{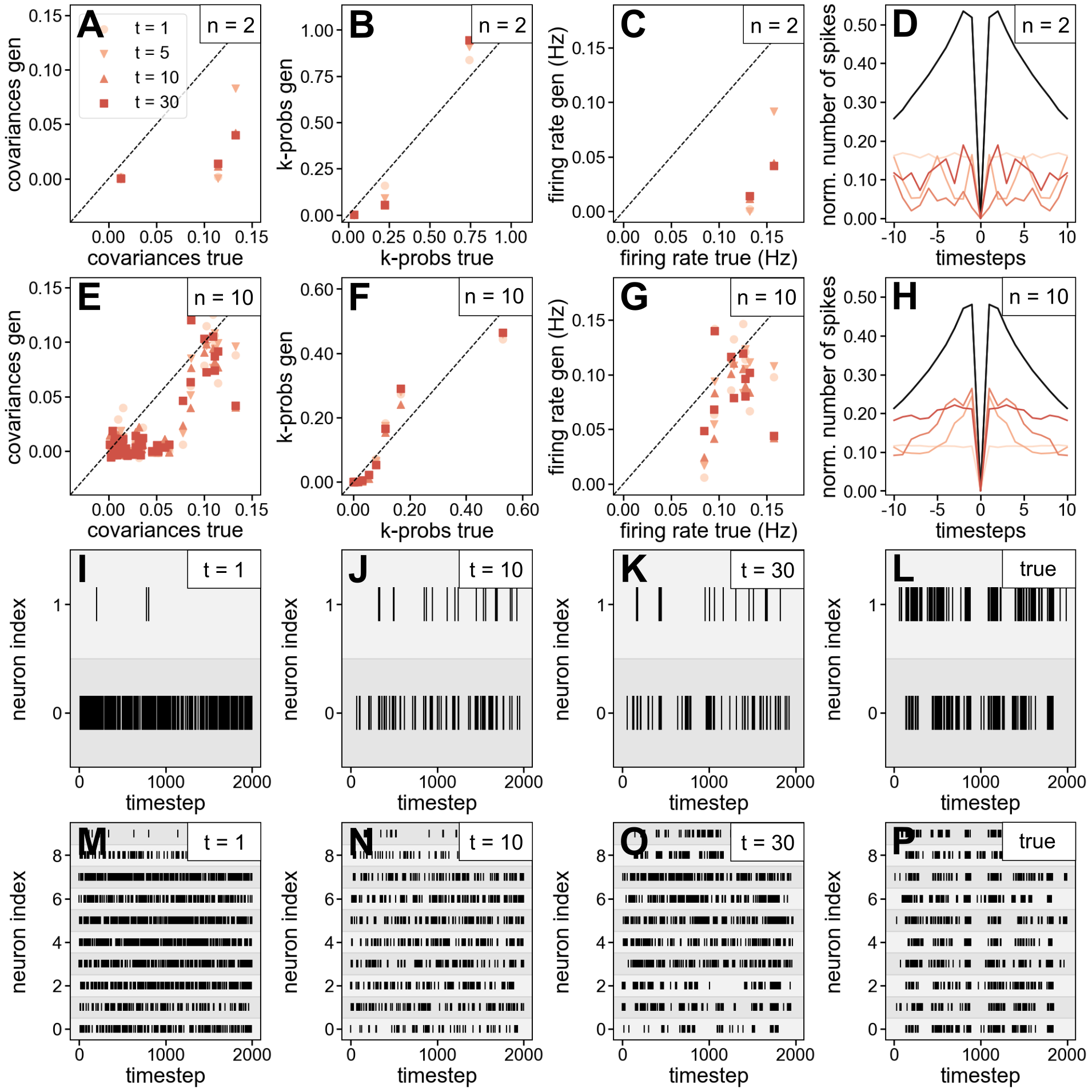} 

	\caption{\textbf{Statistics and generated data for 2 and 10 neurons.}
		(A-H) Statistics for the case of 2 and 10 neurons, with 1, 5, 10, and 30 timesteps represented with different colors in each image. Specifically, (A,E) pairwise covariance, (B,F) k-probability, (C,G) firing rate, and (D,H) autocorrelogram are shown. (I-P) Spike traces for 2 and 10 neurons, for the case of generated data with 1 (I,M), 10 (J,N), and 30 timesteps (K,O), and for real data (L,P).}
	\label{fig:stats} 
\end{figure}

In Fig. \ref{fig:stats} we show further statistics used to assess the quality of the generated data, for $2$ and $10$ neurons and varying the number of time steps, focusing only on the biologically informed \textit{K}-loss from now on. Complete results for all neuron numbers and timesteps, accompanied by a focused comparison between the statistics obtained with two different loss functions, are shown in the Supplementary Text and figures of the Supplementary Material \cite{methods}. In Fig.~\ref{fig:stats} we see that the k-probability (B,F) and the mean firing rate (C,G) are well-fitted by SpiQGAN, while the pairwise covariance (A,E) and the autocorrelogram (D,H) show more discrepancy. In Fig.~\ref{fig:stats}(H), we see that number of generated spikes as a function of simulated time steps gets closer to the real distribution as the number of time steps increases.

 Visual comparison of spike trains generated by SpiQGAN and those from the biological dataset is shown in Fig.~\ref{fig:stats} for $2$ (I-L) and $10$ (M-P) neurons. A visual comparison between the generated spikes and the salamander retina samples shows that for increasing number of time steps the QGAN generated samples start forming bursting clusters, an important feature of the biological dataset. 

Overall, all SpiQGAN iterations we implemented achieved a very good fit of the data while maintaining a low number of parameters, which scale favorably (linearly) in the number of neurons. Specifically, for our model, with $4$ parameterized layers, the total number of trainable parameters is equal to $8$ times the number of neurons per time step. This scaling has the following implication: data generation for neuronal network with dozens of neurons in our implementation uses hundreds of trainable parameters, compared to thousands, or tens of thousands, in the case of the traditional machine learning approaches \cite{MoOnPi2018, LeShlizerman2022}. Moreover, it is clear that models that simulate more neurons presented improved performance, which insinuates that using larger circuits could return even better results. 

\subsection*{Concluding remarks}

We have shown that Quantum Generative Adversarial Networks are able to generate synthetic neuronal activity data that faithfully reproduce both spatial and temporal correlations of the biological dataset. We designed and implemented a resource efficient SpiQGAN that re-uses the same building block across the model. Additionally, we included a biologically informed loss function to take into account statistical properties of the generated samples.

This work lays the foundation for the utilization of quantum learning models beyond quantum science, here in neuroscience modeling. In particular, SpiQGAN opens the possibility of running resource efficient algorithms on quantum computers to beneficially model neuronal activity. With the compact quantum models, the dynamics and interpretation of neuronal activity can be efficiently explored in future work.

\newpage

\clearpage 

%
\bibliography{science_template} 
\bibliographystyle{sciencemag}

%
%
%
%
%
%


\section*{Acknowledgments}
We acknowledge useful discussions with Amira Abbas, Antón Rodriguez-Otero, Thomas Spriggs, Dimphna Meijer, and Geeske van Woerden.

\paragraph*{Funding:}
We acknowledge Kavli Institute of Nanoscience Delft Synergy Grant. This work is part of the project Engineered Topological Quantum Networks (Project No.VI.Veni.212.278) of the research program NWO Talent Programme Veni Science domain 2021 which is financed by the Dutch Research Council (NWO).

\paragraph*{Author contributions:}
EG designed the project with input from VH. VH wrote the code, ran the simulations, analyzed the data and created the images with input from EG. EG and VH co-wrote the paper.

\paragraph*{Competing interests:}
No competing interests.

\paragraph*{Data and materials availability:}

The data used in this work are deposited at Zenodo \cite{dataset_spiqgan}, with accompanying code available at GitLab \cite{gitlab_spiqgan}. 


\subsection*{Supplementary materials}
Materials and Methods\\
Supplementary Text\\
Figs. S1 to S11\\
References \textit{(50-\arabic{enumiv})}\\ 


\newpage


\renewcommand{\thefigure}{S\arabic{figure}}
\renewcommand{\thetable}{S\arabic{table}}
\renewcommand{\theequation}{S\arabic{equation}}
\renewcommand{\thepage}{S\arabic{page}}
\setcounter{figure}{0}
\setcounter{table}{0}
\setcounter{equation}{0}
\setcounter{page}{1} 


\begin{center}
\section*{Supplementary Materials for\\ \scititle}

Vinicius~Hernandes$^{\ast}$,
Eliska~Greplova\\ 
\small$^\ast$Corresponding author. Email: v.hernandes@tudelft.nl\\
\end{center}

\subsubsection*{This PDF file includes:}
Materials and Methods\\
Supplementary Text\\
Figures S1 to S11\\

\newpage


\subsection*{Materials and Methods}

\subsubsection*{Generative Adversarial Networks}

Generative Adversarial Networks (GANs), introduced by Goodfellow et al. \cite{goodfellow2014generative}, are a powerful class of generative models that learn to synthesize data samples by framing the learning process as an adversarial game between two neural networks: a generator and a discriminator. The generator network, \( G \), aims to produce data samples that mimic those drawn from the true data distribution \( P_x \). It takes a noise vector \( z \), sampled from a predefined distribution \( P_z \) (e.g., a Gaussian or uniform distribution), and transforms it into a synthetic data sample, \( G(z) \). Meanwhile, the discriminator network, \( D \), acts as a binary classifier, distinguishing between real samples from the true data distribution and fake samples generated by \( G \).

The training objective is formulated as a minimax game, where the generator tries to minimize the probability of the discriminator correctly identifying generated samples, while the discriminator simultaneously maximizes its ability to correctly classify the samples:

\[
\min_G \max_D E_{x \sim P_x} \left[ \log D(x) \right] + E_{z \sim P_z} \left[ \log (1 - D(G(z))) \right].
\]

Although GANs have achieved remarkable success in various applications (e.g., image synthesis, text generation), they are often plagued by training instabilities such as vanishing gradients and mode collapse \cite{arjovsky2017wasserstein}. These issues arise primarily because the loss function may not provide meaningful gradients when the discriminator is too strong or too weak, leading to poor convergence.

The Wasserstein GAN (WGAN) \cite{arjovsky2017wasserstein} addresses many of the training challenges associated with standard GANs by leveraging the Wasserstein distance (also known as the Earth-Mover distance) to measure the divergence between the true data distribution and the generated data distribution. Unlike the original GAN, the discriminator in WGAN, referred to as a critic, outputs a scalar value instead of a binary classification, quantifying how well the generated samples approximate the real data distribution. The WGAN objective is formulated as:

\[
\min_G \max_{D \in \mathcal{D}} E_{x \sim P_x} [D(x)] - E_{z \sim P_z} [D(G(z))],
\]

where \( \mathcal{D} \) is the set of all 1-Lipschitz functions, enforced through weight clipping or gradient penalties \cite{gulrajani2017improved}. By stabilizing the gradients, WGAN significantly improves convergence behavior, allowing the generator to learn a more accurate representation of the target distribution.

\subsubsection*{Parametrized Quantum Circuits and Quantum GANs}

As quantum computing has advanced, quantum machine learning has emerged as a promising frontier. One key concept is Parametrized Quantum Circuits (PQCs). PQCs consist of a sequence of quantum gates with parameters that are classically optimized. PQCs can encode complex quantum states and can be used to approximate complex distributions.

Building on this foundation, Quantum Generative Adversarial Networks (QGANs) extend the GAN framework into the quantum domain by incorporating quantum components such as quantum generators, quantum discriminators, or both. In QGANs, the generator may be implemented as a PQC, which is trained to generate samples that match the desired distribution. 

\subsubsection*{Implementation of the Quantum Generator and the Classical Critic}

SpiQGAN uses a quantum generator to model the spike activity patterns of retinal ganglion cells. Specifically, we employ a Patch WQGAN approach, where the quantum generator is divided into several sub-generators, each corresponding to a different timestep. Each sub-generator shares the same PQC architecture but has independent trainable parameters, allowing for flexibility in capturing the temporal dynamics of neuronal activity.

The generator begins with a random initial quantum state \( |z\rangle \), which is mapped to the final state \( |g\rangle \) using a data re-uploading scheme \cite{PeCeGi2020}. The quantum circuit consists of five layers, where each layer applies a sequence of parametrized unitaries \( U(\theta_i) \) and noise-encoding unitaries \( U(z) \). The parametrized unitary \( U(\theta_i) \) is implemented using rotation gates around the \( Y \) and \( Z \) axes (\( R_Y \) and \( R_Z \)) and entangling operations (CNOT gates) between adjacent qubits, while the encoding block applies \( R_X \) rotations to each qubit to encode a sampled noise vector. The generator outputs a sequence of activity states for multiple neurons over several timesteps by concatenating the outputs from all sub-generators.

The critic in our QGAN framework is a fully connected classical neural network. The network consists of:
\begin{itemize}
    \item An input layer matching the size of the generated samples,
    \item A hidden layer with 64 neurons using ReLU activation,
    \item An output layer without an activation function, which directly provides a scalar value representing the divergence between the real and generated distributions.
\end{itemize}

\subsubsection*{Training Procedure}

SpiQGAN is trained by optimizing two separate loss functions for the generator and the critic. The critic's loss function aims to maximize the difference between its outputs for real samples \( x \) and generated samples \( G(z) \):

\[
L_C = \frac{1}{2B} \sum_{j} \left( C(G(z_j)) - C(x_j) \right),
\]

whereas the generator's objective is to minimize the critic's evaluation of the generated samples:

\[
L_G = - \frac{1}{B} \sum_j C(G(\boldsymbol{z_j})) - K\left(\sum_i G({z_j})^i-x_j^i\right),
\]

with \( B \) being the batch size, and the term $K(\sum_i G({z_j})^i-x_j^i)$ added to the standard Wasserstein's generator loss function, inspired by Maximum-Entropy models \cite{TkMaMo2013}, corresponding to the difference between the number of spikes in a fake sample and those in a real sample. This loss was named K-loss, and by setting $K=0$ the standard loss is retrieved. Training alternates between two updates of the critic and one update of the generator, ensuring stable convergence. The Adam optimizer is employed with learning rates of 0.05 for the generator and 0.002 for the critic.

\subsubsection*{Dataset and Evaluation Metrics}

The dataset is comprised of neuronal spike activity recorded from retinal ganglion cells in a salamander retina \cite{marre2017multi}. It contains 297 repetitions of a 19-second natural movie, recorded as binary spike events, where $1$ indicates a spike and a $0$ indicates no spike. The goal is to generate synthetic data that replicates these binary spike patterns while maintaining important statistical properties. To evaluate the performance of the QGAN, we used the following statistical metrics:

\begin{itemize}
    \item \textbf{Pairwise Covariance:} Measures the extent to which two neurons fire together. High covariance suggests that the neurons are more likely to spike simultaneously.
    
    \item \textbf{Mean Firing Rate:} The average rate at which a neuron fires spikes over time. This metric helps ensure that the generated data matches the overall activity level of the real data.
    
    \item \textbf{k-Probability:} The probability distribution over the number of spikes (k) in a given time window. Matching this distribution ensures that the generated data captures the variability in spike counts.
    
    \item \textbf{Autocorrelogram:} A measure of the temporal structure of the spike train, representing the correlation of a neuron's spike times with itself over different time lags. This metric is crucial for capturing the temporal dynamics of neuronal activity.
\end{itemize}

To comprehensively assess the model's performance, we conducted experiments varying the number of neurons n=\{2,4,6,8,10\} and timesteps t=\{1,2,5,10,20,30 \}. For small-scale systems, we computed the exact probabilities of all possible spiking states, used to compare the generated and real data distributions using distance measures such as Jensen-Shannon divergence, alongside the metrics listed above.

\subsection*{Supplementary Text}

\begin{figure} 
	\centering
	\includegraphics[width=\textwidth]{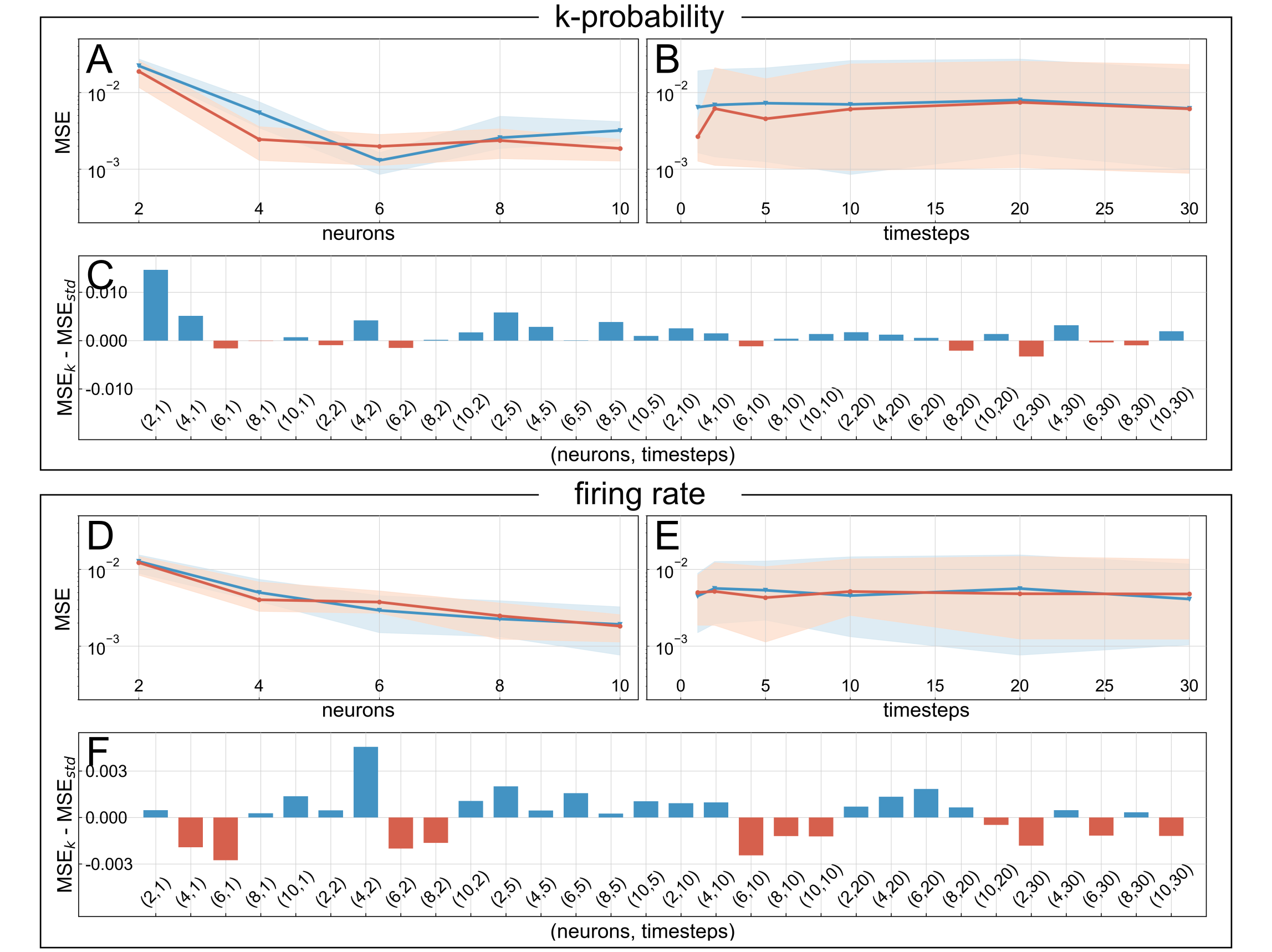} 

	\caption{\textbf{Mean-Square Error of k-probability and firing for models using \textit{K}-loss and standard loss.}
    		(A) Error of k-probability value for \textit{K} (orange) and standard (blue) loss as a function of the number of neurons. (B) Error of k-probability value for \textit{K} and standard loss as a function of the number of timesteps. (C) Difference between the mean-square error of k-probability obtained using the \textit{K}-loss and the standard loss, for all combinations of (neurons, timesteps). (D) Error of firing rate value for \textit{K} (orange) and standard (blue) loss as a function of the number of neurons. (E) Error of firing rate value for \textit{K} and standard loss as a function of the number of timesteps. (F) Difference between the mean-square error of firing rate obtained using the \textit{K}-loss and the standard loss, for all combinations of (neurons, timesteps). }
	\label{fig:loss_comparison} 
\end{figure}

In Fig.~\ref{fig:loss_comparison} we show a comparison between the biological \textit{K}-loss and standard loss. For all combinations of (neurons, timesteps, loss function), we calculate the mean square error between statistics (k-probability and firing rate) obtained with SpiQGAN generated and real samples. Fig.~\ref{fig:loss_comparison}(A) and (D) show the error for both losses for all timesteps as a function of the number of neurons. Fig.~\ref{fig:loss_comparison}(B) and (E) show the error for all neurons as a function of the number of timesteps. The error visibly decreases for increasing number of neurons. Interestingly, the same is not true for increasing number of timesteps, suggesting that even the time correlations (see Fig.~\ref{fig:stats}(H)) are better fitted by using more qubits rather than by increased number of parametrized circuits in the generator. Fig.~\ref{fig:loss_comparison}(C) and (D) shows the difference between the mean-square error obtained using the \textit{K}-loss and the standard loss, for all combinations of (neurons, timesteps). A positive value mean that the error using the \textit{K}-loss is lower, while a negative value means that the error obtained with the standard loss is lower. We see that for most cases, the values are positive, indicating the the \textit{K}-loss achieves a better fit for the majority of models, especially for the k-probability (panel C).

\newpage

\begin{figure} 
	\centering
	\includegraphics[width=\textwidth]{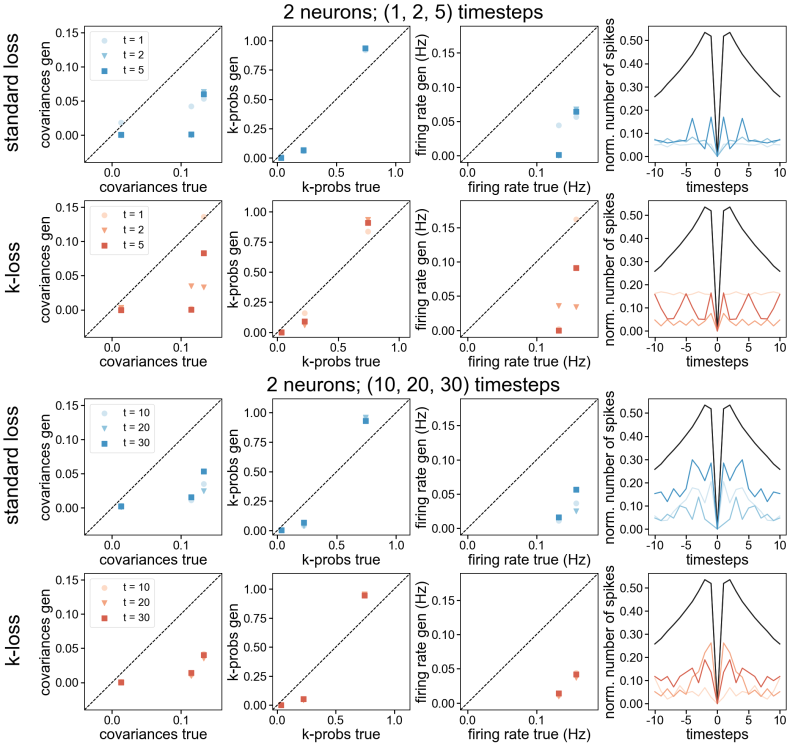} 

	\caption{\textbf{Statistics for 2 neurons.} From left to right: pairwise covariance, k-probability, firing rate, and autocorrelogram. The first row shows the results for the model that uses standard loss, for the case of 1, 2, and 5 timesteps. The second row shows the results for the model that uses K-loss, for the case of 1, 2, and 5 timesteps. Third and fourth rows show the results for 10, 20, and 30 timesteps, for models using standard loss (third) and K-loss (fourth).}
	\label{fig:stats_2} 
\end{figure}

\begin{figure} 
	\centering
	\includegraphics[width=\textwidth]{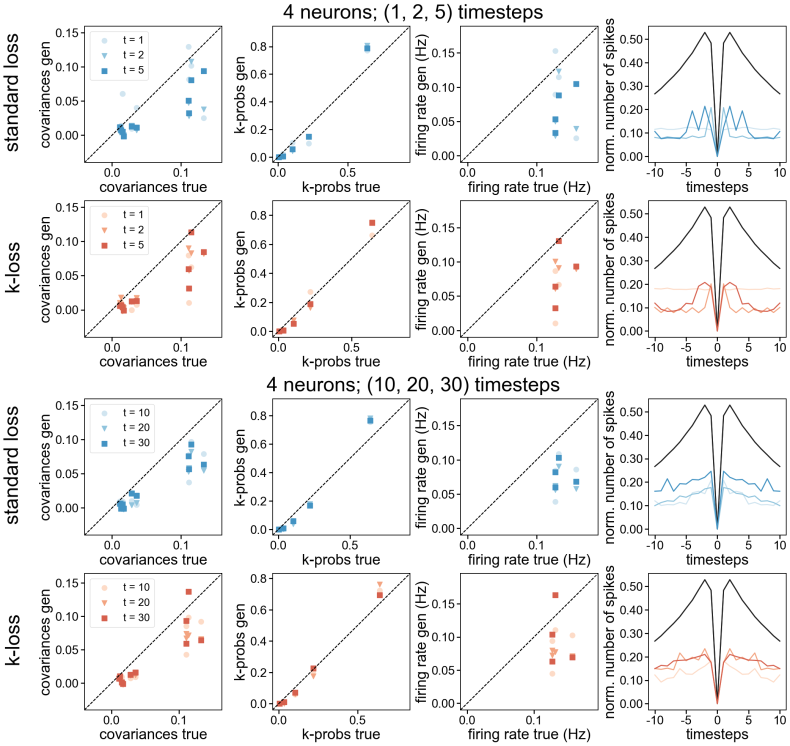} 

	\caption{\textbf{Statistics for 4 neurons.} Same as Fig.~\ref{fig:stats_2}, for 4 neurons.}
	\label{fig:stats_4} 
\end{figure}

\begin{figure} 
	\centering
	\includegraphics[width=\textwidth]{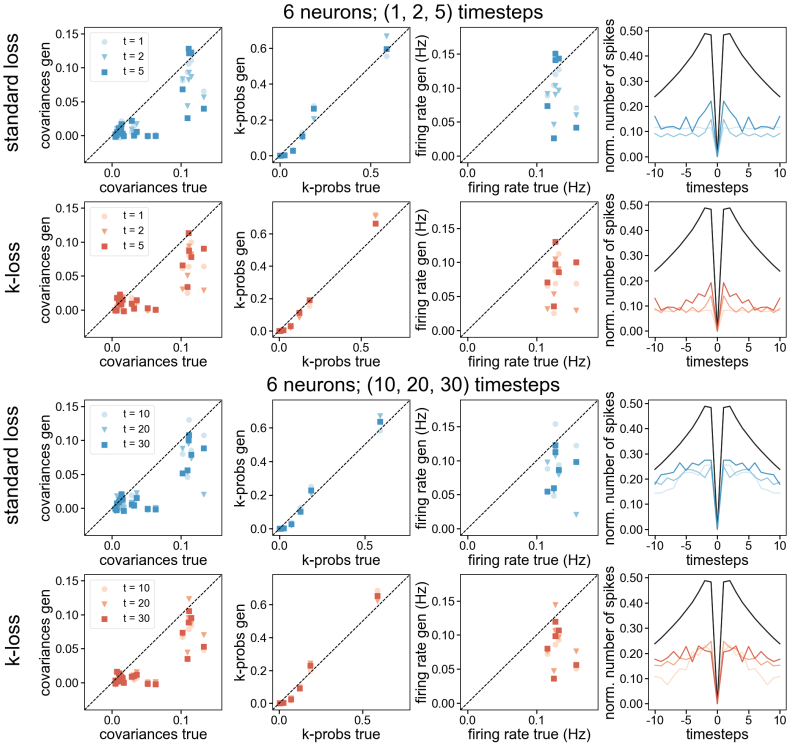} 

	\caption{\textbf{Statistics for 6 neurons.} Same as Fig.~\ref{fig:stats_2}, for 6 neurons.}
	\label{fig:stats_6} 
\end{figure}

\begin{figure} 
	\centering
	\includegraphics[width=\textwidth]{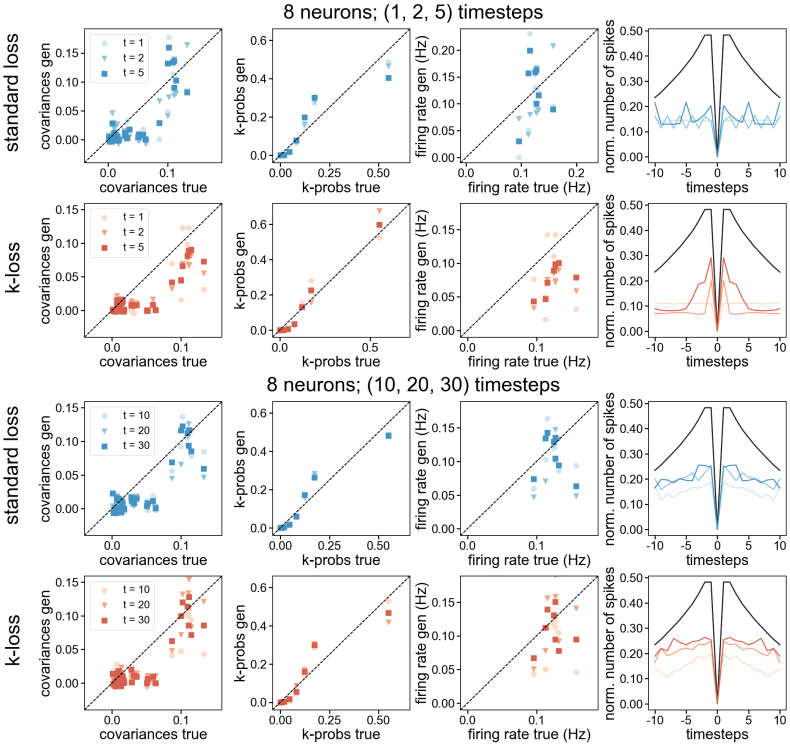} 

	\caption{\textbf{Statistics for 8 neurons.} Same as Fig.~\ref{fig:stats_2}, for 8 neurons.}
	\label{fig:stats_8} 
\end{figure}

\begin{figure} 
	\centering
	\includegraphics[width=\textwidth]{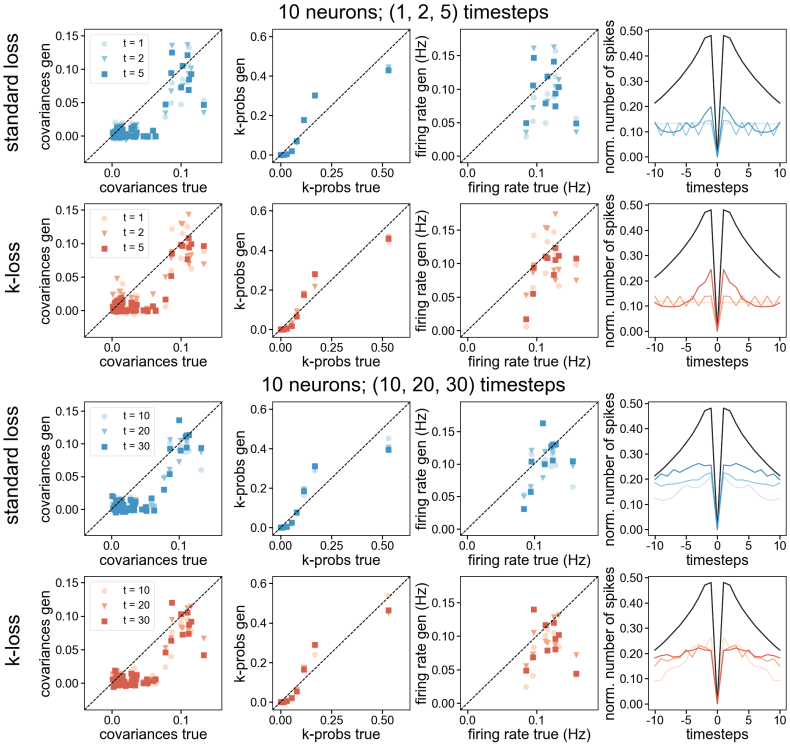} 

	\caption{\textbf{Statistics for 10 neurons.} Same as Fig.~\ref{fig:stats_2}, for 10 neurons.}
	\label{fig:stats_10} 
\end{figure}

\begin{figure} 
	\centering
	\includegraphics[width=\textwidth]{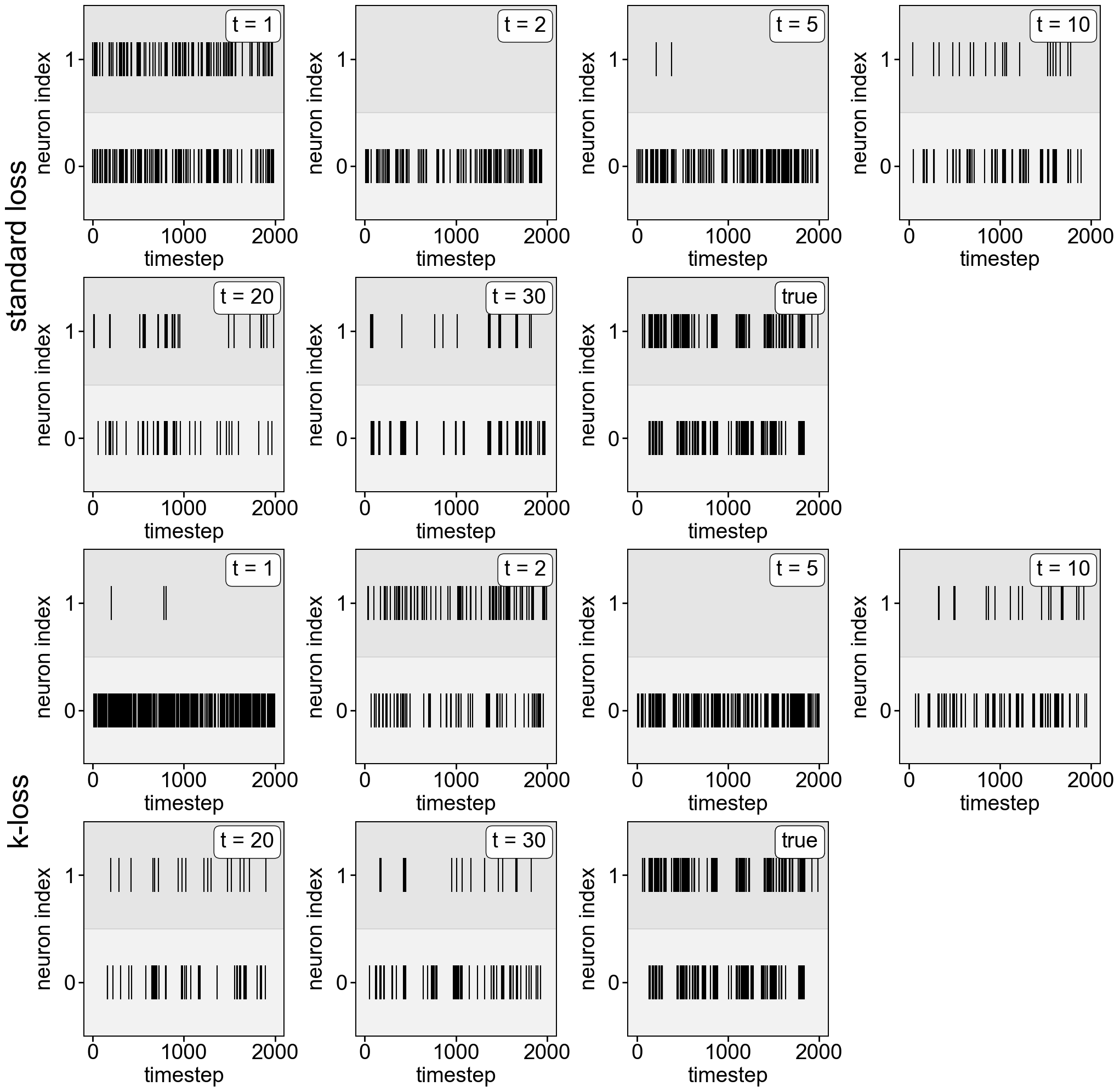} 

	\caption{\textbf{Comparison between generated and real data samples for 2 neurons.}. The first and second rows show spike traces for generated data from trained models with 1, 2, 5, 10, 20, and 30 timesteps, using standard loss, and a spike trace for real dataset. Third and fourth rows show the same as the first two, for the models trained with K-loss.}
	\label{fig:activity_2} 
\end{figure}

\begin{figure} 
	\centering
	\includegraphics[width=\textwidth]{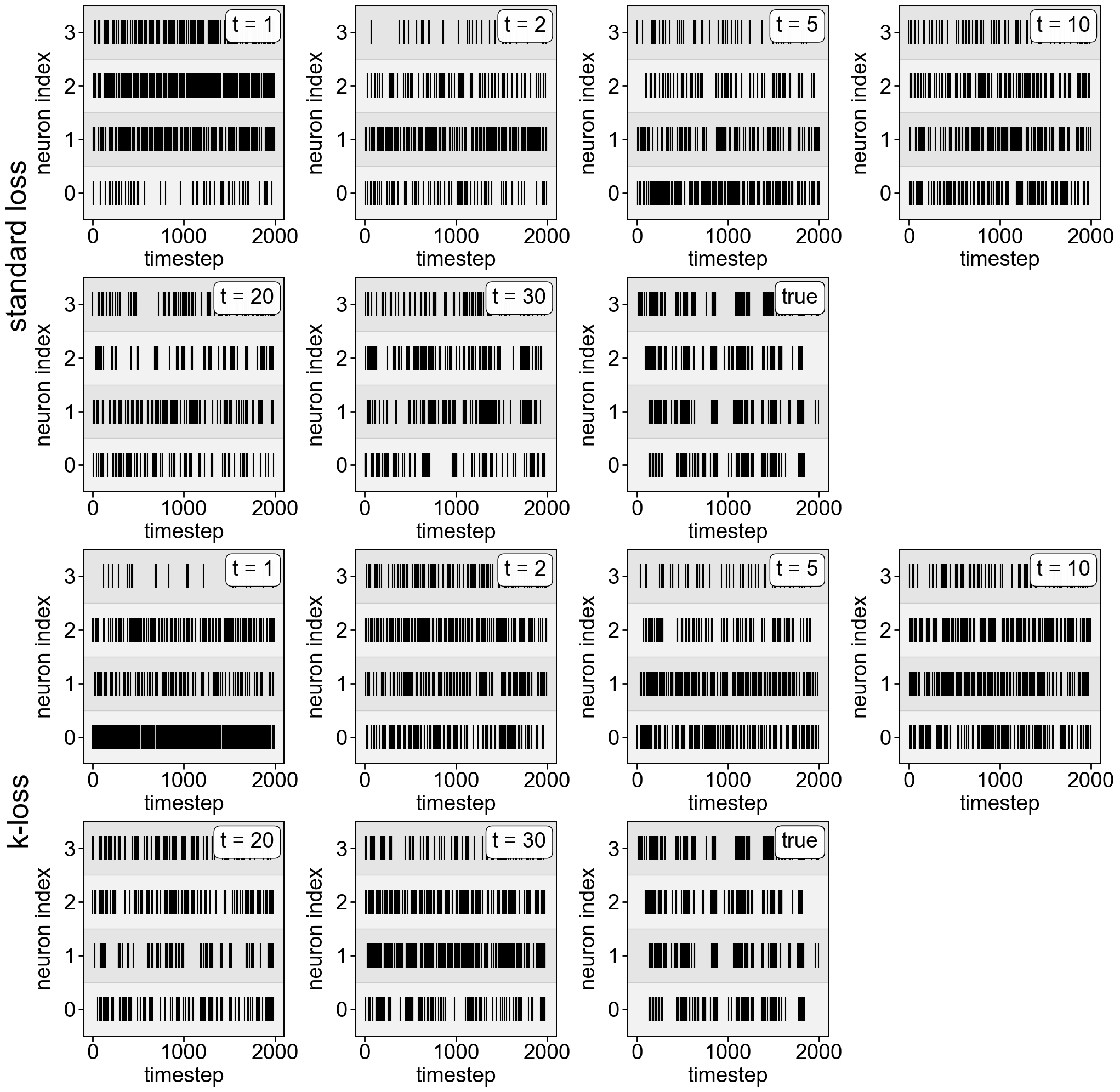} 

	\caption{\textbf{Comparison between generated and real data samples for 4 neurons.}. Same as Fig.~\ref{fig:activity_2}, for 4 neurons.}
	\label{fig:activity_4} 
\end{figure}

\begin{figure} 
	\centering
	\includegraphics[width=\textwidth]{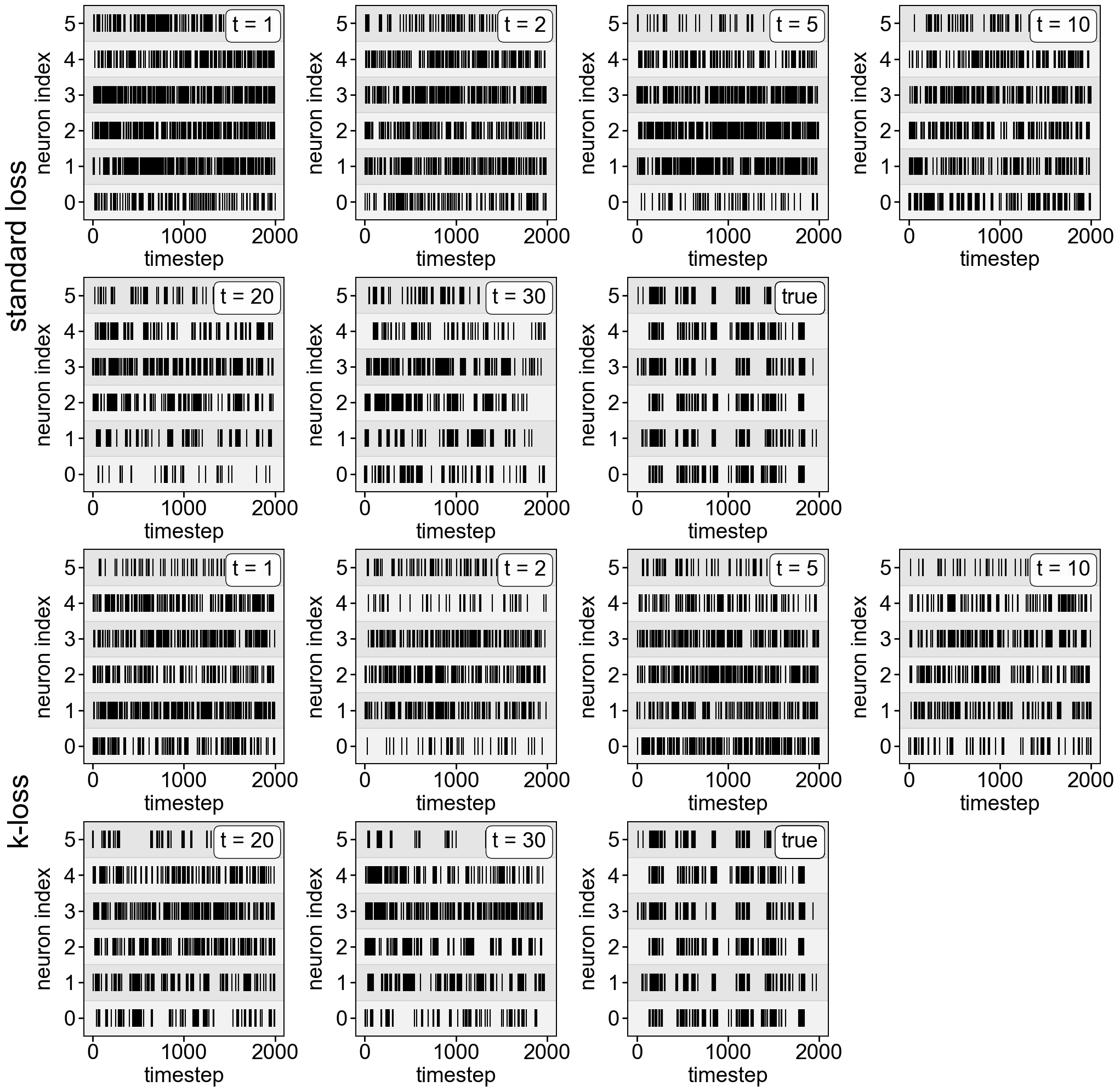} 

	\caption{\textbf{Comparison between generated and real data samples for 6 neurons.}. Same as Fig.~\ref{fig:activity_2}, for 6 neurons.}
	\label{fig:activity_6} 
\end{figure}

\begin{figure} 
	\centering
	\includegraphics[width=\textwidth]{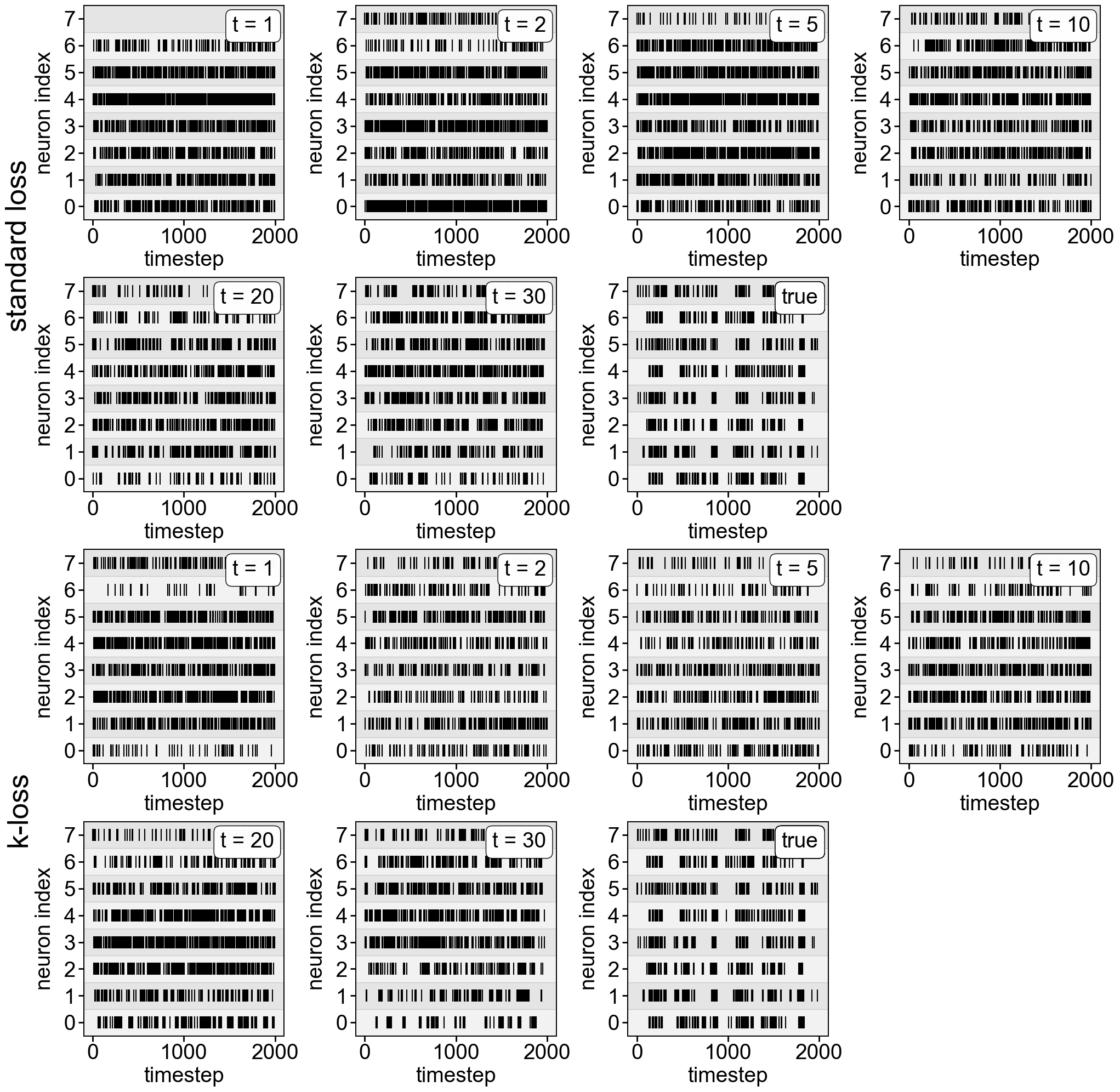} 

	\caption{\textbf{Comparison between generated and real data samples for 8 neurons.}. Same as Fig.~\ref{fig:activity_2}, for 8 neurons.}
	\label{fig:activity_8} 
\end{figure}

\begin{figure} 
	\centering
	\includegraphics[width=\textwidth]{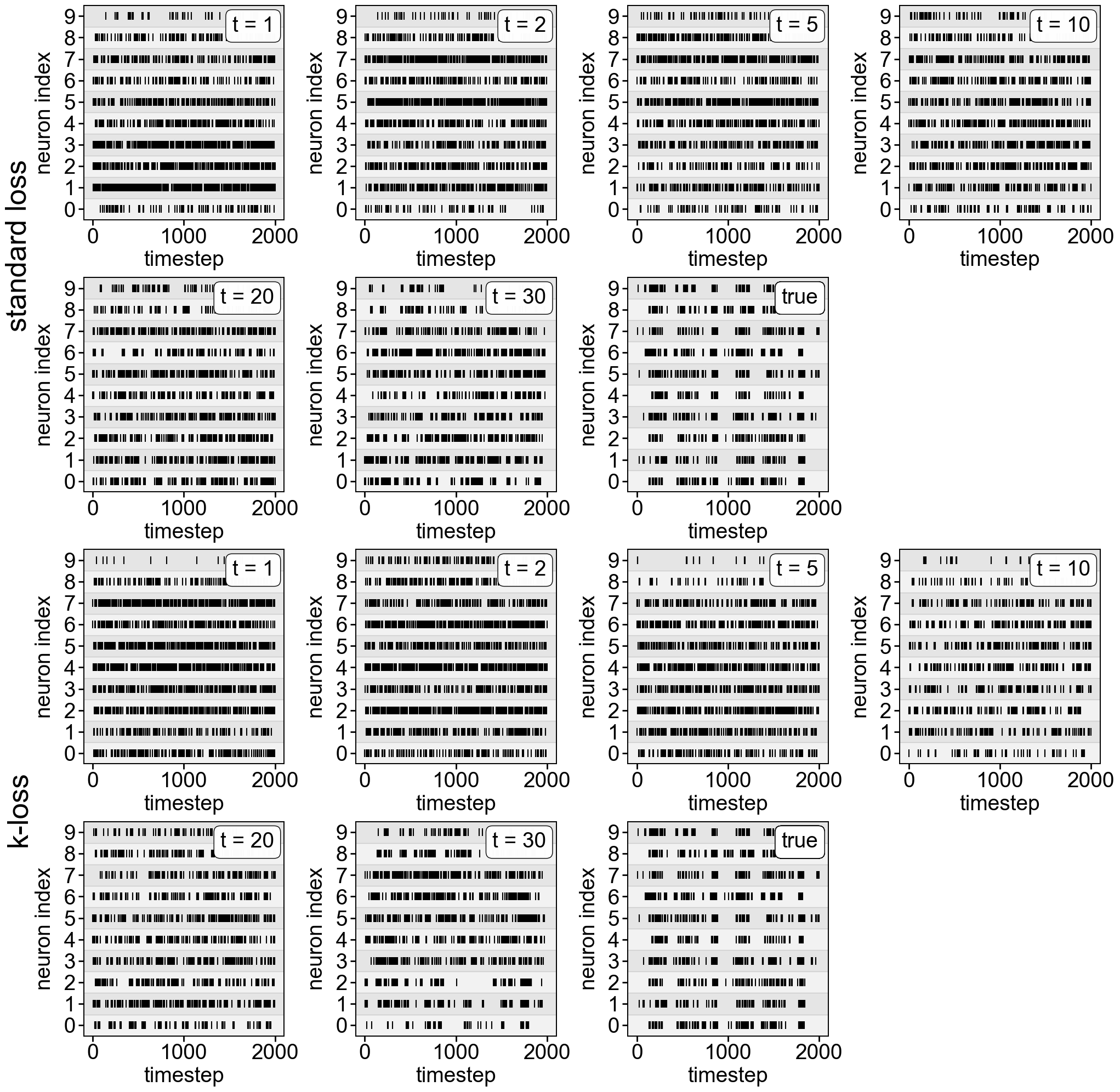} 

	\caption{\textbf{Comparison between generated and real data samples for 10 neurons.}. Same as Fig.~\ref{fig:activity_2}, for 10 neurons.}
	\label{fig:activity_10} 
\end{figure}


\end{document}